\documentclass[twoside]{dis08} 
\usepackage[latin1]{inputenc} 
\usepackage[dvips]{graphicx,epsfig,color} 
\usepackage{wrapfig,rotating} 
\usepackage{amssymb,amsmath,array} 
 
\begin{document} 
\title{W and Anomalous Single Top Production at HERA} 
 
\author{Eram Rizvi\footnote{On behalf of the H1 Collaboration}
%
%
\vspace{.3cm}\\ 
%
Queen Mary, University of London - Dept of Physics \\ 
Mile End Rd, London, E1 4NS - UK
} 
 
\maketitle 
 
\begin{abstract} 

The analysis of $W$ production and the search for anomalous single top
production is performed with the H1 detector at HERA with an
integrated luminosity of 0.5~fb$^{-1}$, consisting of the complete
high energy data from the HERA programme.  Production cross section
measurements of single $W$ production, as well as $W$ polarisation
fractions in events containing isolated leptons and missing transverse
momentum are also presented.  In the context of a search for single
top production an upper limit on the top production cross section
$\sigma_{ep\rightarrow etX} < 0.16$~pb is established at the $95\%$
confidence level, corresponding to an upper bound on the anomalous
magnetic coupling $\kappa_{tu\gamma} < 0.14$.  \end{abstract}
 
\section{Events with Isolated Leptons and ${P}_{T}^{miss}$}
\label{sec:sep}

Events containing a high $P_{T}$ isolated electron or muon and
associated with missing transverse momentum have been observed at HERA
\cite{isoleph1origwpaper,isoleph1newwpaper}.
An excess of HERA~I data events (1994--2000, mostly in $e^{+}p$
collisions) compared to the SM prediction at large hadronic transverse
momentum $P_{T}^{X}$ was reported by the H1 Collaboration
\cite{isoleph1newwpaper}.

\begin{wrapfigure}{r}{0.5\columnwidth}
\centerline{\includegraphics[width=0.45\columnwidth]{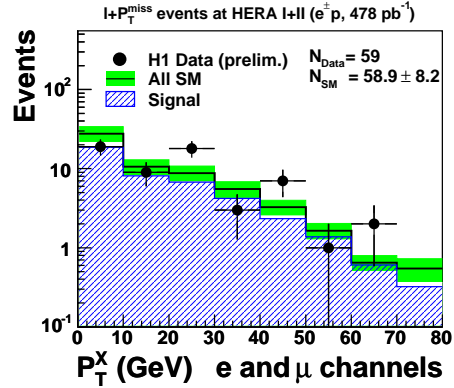}}
  \caption{The $P_{T}^{X}$ 
	distribution of the data (points) compared to the
	SM expectation (open histogram). The signal component of the SM
	expectation is given by the
	hatched histogram. $\rm N_{Data}$ is the total number of data events
	observed, $\rm N_{SM}$ is the total SM expectation. The total error
	on the SM expectation is given by the shaded band.}
\label{fig:isolep1}
\end{wrapfigure}

The main SM contribution is the production of real $W$ bosons via
photoproduction with subsequent leptonic decay $ep\rightarrow
eW^{\pm}$($\rightarrow l\nu$)$X$, where the hadronic system $X$ is
typically of low $P_{T}$.

The event selection employed by the H1 \cite{h1isolepnew} analysis may
be summarised as follows: The identified lepton should have high
transverse momentum $P_{T}^{l} >$~10~GeV, be observed in the central
region of the detector and be isolated with respect to jets and other
tracks in the event.  The event should also contain a large transverse
momentum imbalance, $P_{T}^{miss} >$~12~GeV. Further cuts are then
applied, which are designed to reduce SM background, whilst preserving
a high level of signal purity.

The analysis has recently been performed on the electron and muon
channels using the complete HERA I+II data sets, which corresponds to
478~pb$^{-1}$ \cite{h1isolepnew}. A total of 59 events
are observed in the data, compared to a SM prediction of
58.9~$\pm$~8.2.  For $P_{T}^{X} >$ 25~GeV, a total of 24 events are
observed compared to a SM prediction of 15.8~$\pm$~2.5, of which 21
events are observed in the $e^{+}p$ data compared to a SM prediction
of 8.9~$\pm$~1.5.  The observed data excess in the HERA~I $e^{+}p$
data thus remains at the 3.0$\sigma$ level for the complete H1
$e^{+}p$ dataset.  The results of the analysis are summarised in
Table~$1$.

Figure~\ref{fig:isolep1} shows the $P_{T}^{X}$ distribution of the
$e^{\pm}p$ data for the combined electron and muon channels. The
signal contribution, dominated by real $W$ production, is seen to
dominate the total SM expectation in all data samples. Overall there
is good agreement with the SM expectation. A possible contribution from
anomalous single top production would be expected to contribute at
high $P_{T}^{X}$

\section{Cross Sections and $W$ Polarisation Fractions}

The selection results described in section \ref{sec:sep} are used to
calculate production cross sections for events with an energetic
isolated lepton and missing transverse momentum
($\sigma_{\ensuremath{\ell+{P}_{T}^{miss}}}$) and for single $W$ boson
production ($\sigma_{W})$, for which the branching ratio for leptonic
$W$ decay is taken into account \cite{h1wpol}.  The results are shown
below with statistical (stat) and systematic (sys) uncertainties
compared to the SM prediction, quoted with a theoretical systematic
error (th.sys) of 15\%.

\vspace{-0.2cm}

\begin{table}[h]
\begin{center}
\begin{tabular}{ | l | c @{$\,\pm\,$} c @{$\,(\textrm{stat})\,\pm\,$} c @{\,(sys)\,}| c @{$\,\pm\,$} c @{\,(th.sys)\,}| }
\hline
\multicolumn{1}{|c|}{{ \bf H1}} & \multicolumn{3}{c|}{HERA I+II Data} & \multicolumn{2}{c|}{SM} \\
\hline\hline
{\small $\sigma_{\ensuremath{\ell+{P}_{T}^{miss}}}$} &
{\small 0.24} &
{\small 0.05} &
{\small 0.05} &
{\small 0.26} &
{\small 0.04} \\
{\small $\sigma_{W}$} &
{\small 1.23} &
{\small 0.25} &
{\small 0.22} &
{\small 1.31} &
{\small 0.20} \\
\hline
\end{tabular}
\end{center}
\label{xsec}
\end{table}

\vspace{-0.6cm}

A measurement of the $W$ polarisation fractions is also performed
since new physics may modify the SM polarisation fractions of $W$s
from single top decays~\cite{wpoltheory} and is described in
\cite{h1wpol}.  Additional selection criteria are applied to ensure
good reconstruction of the $W$ and the missing $\nu$. Using a 2D fit,
optimal values of the left-handed ($F_{-}$) and longitudinal ($F_{0}$)
fractions are extracted, as shown in figure \ref{fig:isolep2} (left)
compared to the SM and a FCNC single top model~\cite{anotop}. The data are in
agreement with the SM expectation albeit within large experimental
uncertainties.

\section{Search for Single Top Quark Production}

The excess of events at high $P_{T}^{X}$ may be interpreted in terms
of anomalous single top production via flavour changing neutral
currents with coupling $\kappa_{tu\gamma}$ between $t$ and $u$ quarks
and the exchange photon. Such a search has been reported by H1
previously~\cite{h1top,h1topnew}.

In this analysis, decays of top quarks into a $b$~quark and a $W$~boson
with subsequent decay of the $W$ in the leptonic electron and muon
channels are studied. Therefore a top preselection is applied by
requiring good top mass reconstruction and a lepton charge compatible
top production.

A multivariate analysis is then performed to discriminate top from SM
background (dominated by real $W$~production) using the transverse
momentum of the reconstructed $b$ quark candidate $P_{T}^{b}$, the
reconstructed top mass $M_{\ell\nu b}$, and the $W$ decay angle
$\cos\theta_W^\ell$ calculated as the angle between the lepton
momentum in the $W$ rest frame and the $W$ direction in the top quark
rest frame. A multivariate discriminator is trained using
ANOTOP~\cite{anotop} as the signal model and EPVEC~\cite{epvec} as the
background model. The discriminator is based on a phase space density
estimator using a range search algorithm~\cite{tmva}.

\begin{figure}[b*]
  \includegraphics[height=.47\textwidth]{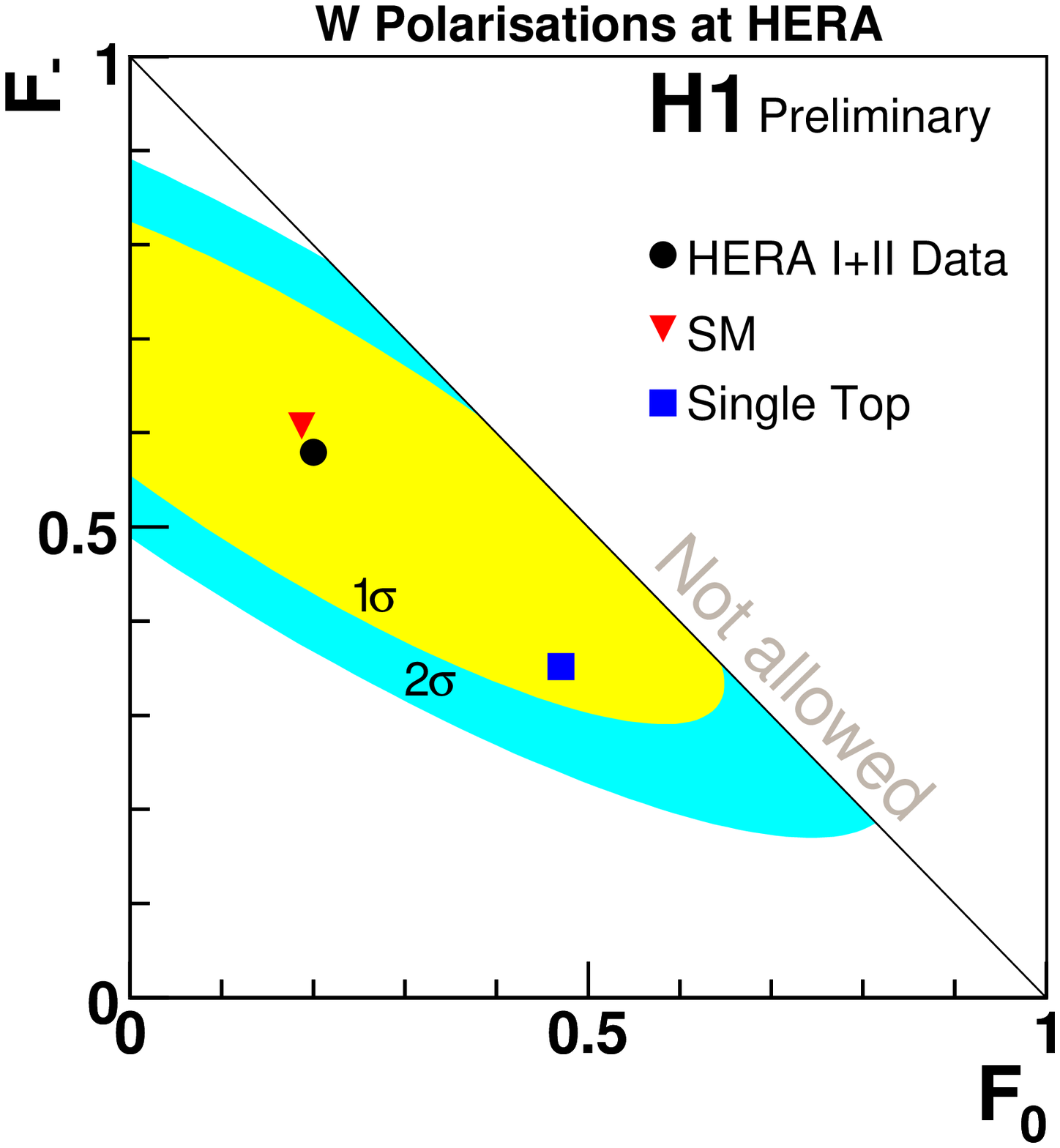}
  \hfill
  \includegraphics[height=.45\textwidth]{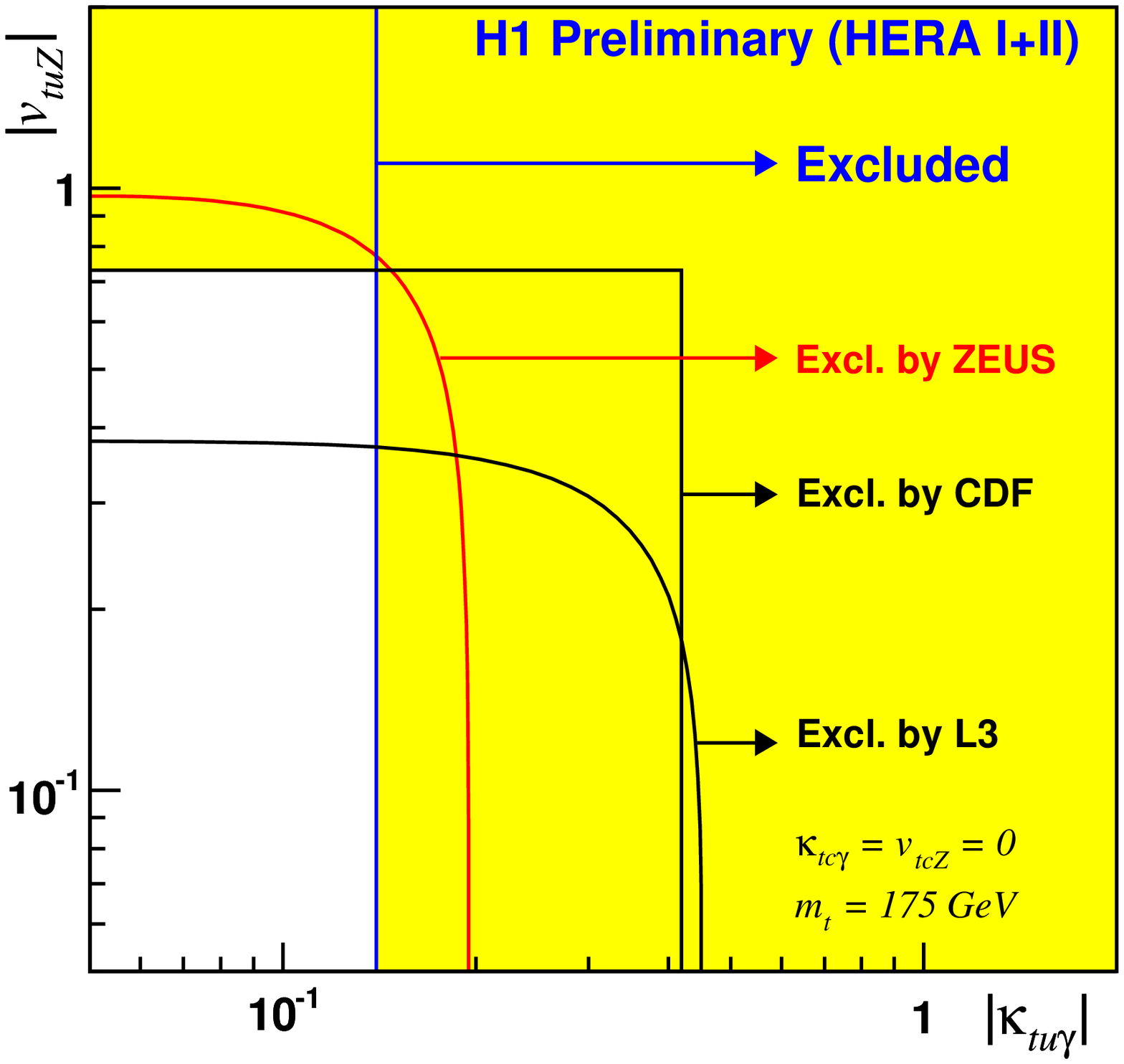}
  \caption{Left: The fit result for the simultaneously extracted left
	handed ($F_{-}$) and longitudinal ($F_{0}$) $W$ boson polarisation
	fractions (point) at 1 and 2$\sigma$ CL (contours). Also shown are
	the values for the SM prediction (triangle) and anomalous single top
	production via FCNC (square). Right: Exclusion limits at the 95\%
	CL in the search for single top production on the anomalous
	$\kappa_{tu\gamma}$ and $v_{tuZ}$ couplings obtained at 
        LEP (L3 experiment \cite{l3top}), the
	TeVatron (CDF experiment \cite{cdftopnew}, the result shown is from~\cite{cdftop}),
	and HERA (H1 \cite{h1topnew}
	and ZEUS \cite{zeustop} experiments). Anomalous couplings of the
	charm quark are neglected $\kappa_{tc\gamma}=v_{tcZ}=0$. Limits
	are shown assuming a top mass $m_t=175$~GeV.}
\label{fig:isolep2}
\end{figure}

The observed data distributions of these quantities agree well with
the SM expectation within the uncertainties. No evidence for single
top production is observed. Using a maximum likelihood method an upper
limit on the anomalous top production cross section of
$\sigma_{ep\rightarrow etX}~<~0.16$~pb is established at 95\% CL.  The
corresponding H1 limit on the coupling $\kappa_{tu\gamma}~<~0.14$ is
shown in figure
\ref{fig:isolep2} (right) and is currently the best limit compared to
those from other colliders~\cite{l3top,cdftopnew}.


\begin{table}[t*]
\begin{center}
 \begin{tabular}{|c|c|c|c|c|}
   \hline
   \multicolumn{2}{|c|}{\large H1 Preliminary} &
   Electron &
   Muon &
   Combined \\
   \multicolumn{2}{|c|}{$l$+$P_{T}^{\rm miss}$ events at} &
   obs./exp. &
   obs./exp. &
   obs./exp. \\
   \multicolumn{2}{|c|}{HERA I+II} &
   {\footnotesize (Signal contribution)} &
   {\footnotesize (Signal contribution)} &
   {\footnotesize (Signal contribution)} \\
   \hline
   \hline
   {\footnotesize $e^{+} p$} &
   {\footnotesize Full Sample} &
   {\footnotesize 26 / 27.3 $\pm$ 3.8 (71\%)}&
   {\footnotesize 15 /  7.2 $\pm$ 1.1 (85\%)}&
   {\footnotesize 41 / 34.5 $\pm$ 4.8 (74\%)}\\
   \cline{2-5}
   {\footnotesize 294 pb$^{-1}$} &
   {\footnotesize $P_{T}^{X}~>25$~GeV} &
   {\footnotesize 11 /  4.7 $\pm$ 0.9 (75\%)}&
   {\footnotesize 10 /  4.2 $\pm$ 0.7 (85\%)}&
   {\footnotesize 21 /  8.9 $\pm$ 1.5 (80\%)}\\
   \hline
   \hline
   {\footnotesize $e^{-} p$} &
   {\footnotesize Full Sample} &
   {\footnotesize 16 / 19.4 $\pm$ 2.7 (65\%)}&
   {\footnotesize  2 /  5.1 $\pm$ 0.7 (78\%)}&
   {\footnotesize 18 / 24.4 $\pm$ 3.4 (68\%)}\\
   \cline{2-5}
   {\footnotesize 184 pb$^{-1}$} &
   {\footnotesize $P_{T}^{X}~>25$~GeV} &
   {\footnotesize  3 /  3.8 $\pm$ 0.6 (61\%)}&
   {\footnotesize  0 /  3.1 $\pm$ 0.5 (74\%)}&
   {\footnotesize  3 /  6.9 $\pm$ 1.0 (67\%)}\\
   \hline
   \hline
   {\footnotesize $e^{\pm} p$} &
   {\footnotesize Full Sample} &
   {\footnotesize 42 / 46.7 $\pm$ 6.5 (69\%)}&
   {\footnotesize 17 / 12.2 $\pm$ 1.8 (82\%)}&
   {\footnotesize 59 / 58.9 $\pm$ 8.2 (72\%)}\\
   \cline{2-5}
   {\footnotesize 478 pb$^{-1}$} &
   {\footnotesize $P_{T}^{X}~>25$~GeV} &
   {\footnotesize 14 /  8.5 $\pm$ 1.5 (68\%)}&
   {\footnotesize 10 /  7.3 $\pm$ 1.2 (79\%)}&
   {\footnotesize 24 / 15.8 $\pm$ 2.5 (73\%)}\\
   \hline
 \end{tabular}
 \caption{Summary of the H1 results of searches for events with isolated
electrons or muons and
       missing transverse momentum for the $e^{+}p$ data (294~pb$^{-1}$), $e^{-}p$ data
       (184~pb$^{-1}$) and the full HERA I+II data set (478~pb$^{-1}$). The results
       are shown for the full selected sample and for the subsample at large $P_{T}^{X}>25$~GeV.
       The number of observed events is compared to the SM prediction. The signal component of the
       SM expectation, dominated by real $W$ production, is given as a percentage in parentheses.
       The quoted errors contain statistical and systematic uncertainties added in quadrature.}
\end{center}
\label{summary}
\end{table}

\end{document}